\documentclass[12pt]{article}      

\title{Relativistic dissipatons in integrable nonlinear Majorana type spinor model}

\author{ Oktay K Pashaev$^1$ and Jyh-Hao Lee$^2$\\ $^1$Department of Mathematics, Izmir Institute of Technology\\ Urla-Izmir, 35430, Turkey \\
$^2$Institute of Mathematics, Academia Sinica, Taipei, Taiwan}

\begin{document}
\newcommand{\be}{\begin{equation}}
\newcommand{\ee}{\end{equation}}
\newcommand{\bea}{\begin{eqnarray}}
\newcommand{\eea}{\end{eqnarray}}
\newcommand{\disp}{\displaystyle}
\newcommand{\la}{\langle}
\newcommand{\ra}{\rangle}

\newtheorem{thm}{Theorem}[subsection]
\newtheorem{cor}[thm]{Corollary}
\newtheorem{lem}[thm]{Lemma}
\newtheorem{prop}[thm]{Proposition}
\newtheorem{definition}[thm]{Definition}
\newtheorem{rem}[thm]{Remark}
\newtheorem{prf}[thm]{Proof}

\maketitle


\begin{abstract}
 By method of moving frame, the relativistic integrable nonlinear model for real, Majorana type spinor fields in 1+1 dimensions is introduced
and gauge equivalence of this model with Papanicolau spin model on one sheet hyperboloid is established. In terms of the so called double numbers, the model is represented also as hyperbolic complex relativistic model, in the form similar to the massive Thirring model. By using Hirota bilinear method,
one dissipaton solution of this model is constructed. 
We calculated first integrals of motion for this dissipaton and show that it represents relativisitc particle with highly nonlinear mass. 
Analyzing resonance conditions for scattering of two relativistic dissipatons, we find a solution describing resonant property of the dissipatons. 
\end{abstract}


\section{Introduction}

The exponentially decaying and finite energy solutions of nonlinear partial differential equations are known as solitons,
and they become indispensable part of integrable systems like KdV and NLS equations, possesing  elastic collision property of soliton interaction.
Another type of solutions exists in dissipative nonlinear equations of reaction-dissusion type, which can grow or decay exponentially.
The finite energy solutions of this kind was called the "dissipatons" \cite{Pashaev1997}, as solutions of integrable system of reaction-diffusion equations,
which was introduced for description of low dimensional gravity model of constant curvature, the Jackiw-Teitelboim (JT) gravity \cite{MPS1997}. This 
system admits arbitrary $N$-dissipaton solutions, showing the resonant property under collision of dissipatons, by creating long time living resonances.
Dissipaton resonances were related with black hole solutions of JT gravity and main characteristic of black holes as existence of the event horizon,
intrinsically connected with resonant property of dissipatons  \cite{MPS1998}. Reformulation of the reaction-diffusion model as the nonlinear Schr\"odinger equation with 
de Broglie-Bohm quantum potential term was proposed 20 years ago in our paper  \cite{PL2002}, as a result of our joint work between 1997-2000 in Academia Sinica, Taipei, Taiwan. The equation
was coined as the Resonant Nonlinear Schr\"odinger equation (RNLS), since envelope solitons of this equation interact by creating resonant soliton states \cite{PLANZIAM2002}.
Then the model was studied intensively in different aspects of phyisical and mathematical characters. The RNLS equation as descriptive in cold plasma physics was proposed in \cite{LPRS2007}, 
\cite{LP2007}, and applied to kinetic of soliton gas in \cite{CER2021}, \cite{El2021}, being subject of experiments in \cite{Redor2019}. Since RNLS appeared in both, the gravity theory and in 
plasma physics, it inspired also research on studying the analog gravity with
 black hole type configurations in plasma physics \cite{Williams}. Another application is 
related with cappilary models of Korteweg types \cite{RogersSchief1999}, \cite{RogersSchief2014}, \cite{Rogers2019}. From mathematical point of view, wide classes of solutions were derived for RNLS 
and its different modifications to  higher dimensions and variable coefficients, see for example
\cite{Tozar2021}, \cite{Li2013}, for symmetry analysis, loop algebraic structure and integrability see \cite{Alfinito1998} - \cite{Demskoi2014}. In addition, the mapping of the RNLS 
hierarchy, the second and the third flow  to KP-II equation 
\cite{PashaevFrancisco}, established link between dissipaton and envelope soliton resonances with planar solitons of KP-II, creating 
the web type structure in shallow water \cite{Kodama2004}, \cite{CK2008}. Several modifications of RNLS models, as the derivative RNLS \cite{PLANZIAM2002}, \cite{LLP2000}, \cite{LeePashaev2005}, 
modified RNLS \cite{LPModifiedNLS},
generic RNLS  \cite{PLR2008}  and related generalized equations  \cite{PL2012}, \cite{LP2014}, \cite{RogersPashaev2011}, \cite{NabelekZakharov2020}, \cite{Nabelek2021},   were studied. However, all these developments are related with non-relativistic dissipatons and envelope solitons, so it is not clear if exist relativistic
nonlinear equations with dissipaton solutions and resonant property of their mutual interaction. 

The goal of the present paper is to show that there exist such model and it admits relativistic dissipaton solution with resonant character of interaction.
The model is derived from $\sigma$ model in constant external field on the one sheet hyperboloid $SO(2,1)/O(1,1)$ and represents relativistic, real-valued spinor fields,
satisfying 1+1 dimensional Dirac type equation with Thirring type nonlinearity. The real solutions of Dirac type equation are knows as Majorana spinors or fermions \cite{Jackiw},
and these solutions appeared recently in condensed matter physics for modeling topological superconductor systems \cite{Sato}. Another reason to study Majorana fermions is connected with problem of
 neutrino
mass, which requires, instead of the Weyl massless  equation to use as descriptive the massive Majorana equation. The model we propose here is integrable relativistic equation 
with nonlinearity of the Thirring type. The Thirring model is one of the best known relativistic nonlinear systems, for which many properties like   integrability \cite{Mikh}, solvability \cite{Lee},
inverse scattering transform \cite{KuzMikh}, \cite{KaupNewell}, \cite{KL1996} and bilinear representation \cite{Chen} were studied. The model we propose here, is also integrable system with Lax pair and bilinear representation. We show that in terms of hyperbolic complex numbers it can be even rewritten in form very similar to the Thirring model. 
But in contrast to Thirring model it admits dissipaton solutions with relativisitc dispersion and leading to resonant character of dissipaton interactions.

The paper is organized as follows. In Section 2 we briefly review a relation between JT gravity model and flat connection BF gauge theory on the one sheet hyperboloid. The gauge fixing conditions in the
theory
are determined by non-linear $\sigma$ models on  $SO(2,1)/O(1,1)$ space and solutions of the models in tangent space give the Riemannian metric tensor of JT gravity.
In Section 3 we introduce nonlinear $\sigma$ model in this space with constant external field, which is the non-compact version of the model introduced by Papanicolau \cite{Papa}.
In tangent space this give us the real spinor relativistic nonlinear model and corresponding zero curvature representation. In Section 4 we derive bilinear form for this model and find one dissipaton
solution. By calculating first three integrals of motion in Section 5 we show that our dissipaton represents relativistic particle type object with highly
nonlinear mass. The analysis of resonance conditions for fusion and fission of scattering dissipatons with relativistic dispersion shows that dissipatons in the model can interact in the resonant way.

\section{{SL(2,R) gauge group and dissipative equations}}

\subsection{Gauge theory of Jackiw-Teitelboim Gravity}

A nontrivial gravity model in 1+1 dimensions
introduced by Jackiw and
Teitelboim is described by the action with Lagrange density
\begin{equation} L = {\sqrt -g}\eta (R - \Lambda), \end{equation}
where $\eta$ is an additional gravitational variable called a world
scalar Lagrange multiplier field, $R$ is the Riemann scalar and
$\Lambda$ is a cosmological constant.
This model can be reformulated
as the $BF$ topological gauge theory with the three-parameter
$SO(2,1)$ de Sitter or anti-de Sitter groups,
$$S = \int_{\Sigma} Tr (\Phi F) =
\Phi^{a} (de^{a} + \epsilon^{a}_{b}\omega e^{b})
+ \Phi(d\omega + {\Lambda \over 4}e^{a}\epsilon_{ab}e^{b}),$$
where $\Phi$ are zero-form Lagrange multipliers, $e^{a}_{\mu}$ is the Zweibein and $\omega$ is the spin connection.
 Variation of
these fields
produces field equations for
the curvature two-form
$$F = dA + A^{2} = (de^{a} + \epsilon^{a}_{b}\omega e^{b})P_{a}
+ (d\omega + {\Lambda \over 4}e^{a}\epsilon_{ab}e^{b})J =0,$$
giving the torsionless and curvature conditions  
$$de^{a} + \epsilon^{a}_{b}\omega e^{b} = 0,\,\,\,\,
d\omega + {\Lambda \over 4}e^{a}\epsilon_{ab}e^{b} =0,$$
equivalent to the Jackiw-Teitelboim model (2.1).
The Riemann metric tensor $g_{\mu\nu}$ can be recovered from Zweibein fields
according to the relation
$$g_{\mu\nu} = e^{a}_{\mu}e^{b}_{\nu}\eta_{ab} = -{4 \over\Lambda}(q^{+}_{\mu}q^{-}_{\nu}
+ q^{+}_{\nu}q^{-}_{\mu}).$$
where $\eta_{ab} = diag(1, -1)$ is a flat tangent space metric,
and
spin connection and Zweibeins are
$$V_{\mu} = 2\omega_{\mu},$$
$$ q^{\pm}_{\mu} \equiv u_{\mu} \pm w_{\mu} = {1 \over 2}
{\sqrt {-{\Lambda\over 2}}}(e^{0}_{\mu} \pm e^{1}_{\mu}) \equiv {1 \over 2}
{\sqrt {-{\Lambda\over 2}}}e^{\pm}_{\mu}.$$
These relations give us possibility of gravitational interpretation
for our models by considering different
integrable nonlinear $\sigma$ models on the one sheet hyperboloid
$SL(2,R)/O(1,1)$. Reformulated in the tangent space,
they represent the gauge fixing conditions for the BF topological
gauge theory. The resulting equations with global
 O(1,1) gauge symmetry group represent nonlinear dissipative equations in real variables.

\subsection{Moving Frame for Poincare Gauge Group in 1+1 dimension}

Here we briefly review the gauge theoretical treatment of
noncompact SO(2,1) $\sigma$ models
with Abelian $O(1,1)$ subgroup \cite{Pashaev1997}, \cite{MPS1998}. These models
are relevant to the 1+1 dimensional Jackiw-Teitelboim model,
where the subgroup plays the role of 
Lorentz transformation in the tangent plane.
From another side, they lead to the dissipative nonlinear
systems, like the reaction-diffusion system. The relation between
these two, at first sight looking different fields is instructive.

We consider the group $SL(2,R)$ with element $g$, generated by
$\tau_{i}$ $(i=1,2,3)$, satisfying
\begin{equation}
\tau_{i}\tau_{j} = h_{ij} + ic_{ijk}\tau_{k},
\end{equation}
where $h_{ij}$ and $c_{ijk}$ are the Killing metric
and structure constants of
$SL(2,R)$.
Explicit realization in terms of Pauli matrices $\sigma_{i}$ is
$\tau_{1} = -i\sigma_{1}, \tau_{2} = \sigma_{2}, \tau_{3} = -i\sigma_{3}.$
We define an orthonormal trihedral set of unit vectors
${\bf n}_{i}$ 
in the adjoint
representation of $SL(2,R)$,
\begin{equation}
({\bf n}_{i},\tau) =
{\bf n}^{k}_{i}\tau_{k} = h_{kl}{\bf n}^{k}_{i}\tau^{l} =
g \tau_{i} g^{-1}.\end{equation}
The inner and
cross products between
three-vectors are defined as
\begin{eqnarray}
({\bf n}_{i},{\bf n}_{j}) &=& h_{ij} ,\\
{\bf n}_{i}\wedge  {\bf n}_{j} &= & c_{ijk} {\bf n}_{k},
\end{eqnarray}
where
$h_{ij} = diag(-1,1,-1)$
and
\begin{equation}c_{ijk} = {1 \over 2} tr(\tau_{i}\tau_{j}\tau_{k})h_{kk},\end{equation}
or explicitly in terms of the antisymmetric constant tensor
$\epsilon_{ijk}$,
\begin{equation} c_{ijk} = - \epsilon_{ijk}h_{kk}.\end{equation}
Let ${\bf n}_{i}= {\bf n}_{i}(x^0, x^1)$ are smooth
vector fields that define at each space-time coordinates $ (x^{0},x^{1})$ 
a moving frame (orthonormal basis).
By the right-invariant chiral current,
\begin{equation}
J_{\mu } = g^{-1}\partial _{\mu }g,\,\,\,\,\mu =0,1, \label{J}\end{equation}
the moving frame rotates according to equation
\begin{equation}
\partial _{\mu }{\bf n}_{i}= (J^{R}_{\mu })^{(ad)}_{ik}{\bf n}_{k}.
\end{equation}
We
decompose  matrix $J_{\mu }$ to diagonal and off diagonal
parts,
$$
J_{\mu } = J^{(0)}_{\mu }+ J^{(1)}_{\mu },
$$
parametrized in the following form
\begin{eqnarray}
J^{(0)}_{\mu } &=& {i \over 4} \tau _{3}V_{\mu } ,\\
J^{(1)}_{\mu }& =& iu_{\mu}\tau_{1} - iw_{\mu}\tau_{2} =
\left(\matrix{0&u_{\mu} - w_{\mu}\cr u_{\mu} + w_{\mu}&0 \cr}\right).
\end{eqnarray}
Vector ${\bf s \equiv  n}_{3}$ satisfies the constraint
${\bf s}^2 =
({\bf s}(x), {\bf s}(x)) = - s_{1}^{2} + s_{2}^{2} - s_{3}^{2} =-1 $
and belongs to the one sheet hyperboloid $S^{1,1} \sim SL(2,R)/O(1,1)$.
The real fields $V_{\mu }$, $u_{\mu}$ and $w_{\mu }$ are recovered by projections,
$$
V_{\mu } = 2({\bf n}_{2}, \partial _{\mu }{\bf n}_{1}),\,\,\,\,
w_{\mu } =
+{1 \over 2}({\bf s, \partial }_{\mu }{\bf n}_{1}),\,\,\,\, u_{\mu } =
 {1 \over 2}({\bf s, \partial }_{\mu }{\bf n}_{2}).
\eqno(1.20)$$
In the
light-cone basis,
\begin{equation}
{\bf n}_{+}= {\bf n}_{1} + {\bf n}_{2} , \,\,\,\,{\bf n}_{-}=
{\bf n}_{1} - {\bf n}_{2},
\end{equation}
 satisfying following relations
\begin{eqnarray}
({\bf n}_{+}, {\bf n}_{+}) &=& 0 = ({\bf n}_{-}, {\bf n}_{-}), \,\,\,\,\,
({\bf n}_{+}, {\bf n}_{-}) = - 2,\\
{\bf n}_{+}\wedge  {\bf s} &=& + {\bf n}_{+} ,\,\,\,\,
{\bf n}_{-}\wedge  {\bf s} = - {\bf n}_{-} ,\,\,\,\,\,
{\bf n}_{-}\wedge  {\bf n}_{+} = 2{\bf s}. 
\end{eqnarray}
we define real fields 
\begin{equation}
q^{+}_{\mu } = u_\mu + w_\mu = + {1 \over 2} ({\bf s}, \partial_{\mu}{\bf n}_{+}),\,\,\,\,
{q}^{-}_{\mu } = u_\mu - w_\mu =
- {1 \over 2} ({\bf s}, \partial_{\mu}{\bf n}_{-}) .
\end{equation}
 In terms of these variables the moving frame equations become
\begin{eqnarray}
D^{-}_{\mu }{\bf n}_{+} &=& -2 q^{+}_{\mu } {\bf s} , \label{mf1}\\
D^{+}_{\mu }{\bf n}_{-} &=& +2 q^{-}_{\mu } {\bf s} \label{mf2},\\
\partial _{\mu }{\bf s} &=& q^{+}_{\mu } {\bf n}_{-} -
q^{-}_{\mu } {\bf n}_{+} \label{mf3} ,
\end{eqnarray}
where $D^{\pm}_{\mu } \equiv  \partial _{\mu } \pm (1/2) V_{\mu }$
is the covariant derivative.
This   form   is   explicitly invariant under the local $O(1,1)$
gauge transformations,
\begin{equation}
{\bf s} \rightarrow {\bf s} , \,\,\,\,\,{\bf n}_{+}\rightarrow e^{+\alpha }
{\bf n}_{+} , \,\,\,\,\,{\bf n}_{-} \rightarrow e^{-\alpha } {\bf n}_{-} ,
\end{equation}
 which are just the Lorentz boost rotations in the tangent to the
vector {\bf s} plane. Finally, consistency conditions of system (\ref{mf1}), (\ref{mf2}), (\ref{mf3}),
are equations for fields $V_{\mu }$ and $q_{\mu }$,
\begin{eqnarray}
D^{-}_{\mu }q^{+}_{\nu }&=& D^{-}_{\nu }q^{+}_{\mu },\label{ZC1} \\
D^{+}_{\mu }q^{-}_{\nu }&=& D^{+}_{\nu }q^{-}_{\mu },\label{ZC2} \\
\partial_{\mu}V_{\nu} - \partial_{\nu}V_{\mu}& =& 4
(q^{+}_{\mu }q^{-}_{\nu } -
q^{+}_{\nu }q^{-}_{\mu }) \label{ZC3} , 
\end{eqnarray}
representing the zero-curvature conditions for current (\ref{J}), parametrized now as 
\begin{equation} J_{\mu} = {i\over 4}V_{\mu}\tau_{3} +
\left(\matrix{0 & q^{-}_{\mu}  \cr q^{+}_{\mu} & 0 \cr   } \right).\label{JMU}
\end{equation}

\subsection{Resonant NLS equation} 

For the Heisenberg model on one sheet hyperboloid ${\bf s} \in SO(2,1)/O(1,1)$,
\begin{equation}
\partial_0 {\bf s} = {\bf s} \wedge \partial^2_1 {\bf s} \label{HM}
\end{equation}
this method produces integrable system of reaction-diffusion equations 
\begin{equation}
\mp \partial_0 q^{\pm} + \partial^2_1 q^{\pm} - 2 q^+ q^- q^{\pm} =0 \label{RD}
\end{equation}
for pair of real functions \cite{Pashaev1997}. It can be transformed to the called Resonant NLS equation \cite{PL2002}, 
which includes the de Broigle-Bohm quantum potential
\begin{equation}
i \psi_t + \psi_{xx} + |\psi|^2 \psi = 2 \frac{|\psi|_{xx}}{|\psi|} \psi.
\end{equation}
This equation admits envelope solitons with resonant interaction and 
besides application for JT gravity model \cite{Pashaev1997}, \cite{MPS1997}, \cite{MPS1998}, it was derived also in several physical models as cold plasma physics \cite{LPRS2007}, \cite{LP2007} and 
capillary models\cite{RogersSchief1999}, \cite{RogersSchief2014}, \cite{Rogers2019}. Existence of regular dissipaton solutions for system (\ref{RD}) is crucial for resonant soliton interactions in RNLS.

\subsection{Relativistic models}

Several relativistic models can be derived from the topological magnetic fluid model, proposed in \cite{MPS1994}, which can be bilinearized in arbitrary number of dimensions. 
The model can be formulated also for noncompact spin ${\bf s} \in SO(2,1)/O(1,1)$, as the system of Landau-Lifshitz equation in moving frame with velocity $v^\mu$, and
relation between vorticity of the flow and topological spin density,
\begin{eqnarray}
\partial_0 {\bf s} + v^\mu \partial_\mu {\bf s} = {\bf s} \wedge \partial^\mu \partial_\mu {\bf s}, \label{TM1}\\
\partial_\mu v_\nu - \partial_\nu v_\mu = 2 {\bf s} \cdot (\partial_\mu {\bf s} \wedge \partial_\nu {\bf s}).
\end{eqnarray}
The Heisenberg model on one sheet hyperboloid (\ref{HM}) is particular reduction of this system 
in 1+1 dimensions, with vanishing velocity field $v^\mu =0$. Due to resonant character of soliton interactions in the last spin model (see \cite{PL2002}) and corresponding reaction-diffusion and RNLS equations, having several 
applications to non-relativistic physical systems, it is interesting problem to construct relativistic invariant systems, admitting dissipaton solutions  with resonant scattering properties.

\subsubsection{Self-dual $\sigma$ model} For self-dual $\sigma$ model
\begin{equation}
\partial_0 {\bf s} = {\bf s} \wedge \partial_1 {\bf s}
\end{equation}
the corresponding equations in tangent space \cite{MPS1998} are given by non-linear and non-local relativistic system of equations for real (Majorana type) fields 
\begin{eqnarray}
-\partial_- q^+_+ + q^+_+ \int^x q^+_+ q^-_- dx' = 0, \\
\partial_+ q^-_- + q^-_- \int^x q^+_+ q^-_- dx' = 0.
\end{eqnarray}
This system can be solved by substitution 
\begin{equation} 
q^+_+ = e^{R+S},\,\,\,q^-_- = e^{R -S},
\end{equation}
giving the Liouville equation 
\begin{equation}
\partial^2_0 R - \partial^2_1 R = e^{2 R}. \label{Liouville}
\end{equation}
The general solution of Liouville equation is given in terms of arbitrary real functions
$A(s)$ and $B(s)$ of one variable $s$,
\begin{equation}
R(x^0, x^1)= \frac{1}{2} \ln \frac{A'(x^0 + x^1) B'(x^0 - x^1)}{(A(x^0 + x^1) + B(x^0 - x^1))^2}.
\end{equation}
Then, for any solution of this equation, function $S$ can be obtained by integration
of linear system 
\begin{equation}
\partial_0 S = \partial_1 R + \int^x e^{2 R} dx',\,\,\,\,\partial_1 S = \partial_0 R.
\end{equation}
Compatibility condition for the last one is just the Liouville equation (\ref{Liouville}).
\subsubsection{Nonlinear $\sigma$ model}

Another relativistic model  is associated with nonlinear $\sigma$ model 
\begin{equation}
\partial_+ \partial_- {\bf s} -  (\partial_+ {\bf s} \cdot \partial_- {\bf s}) {\bf s} =0.
\end{equation}
The tangent space representation of this equation leads to a more general nonlinear relativistic model \cite{MPS1998},
\begin{eqnarray}
-\partial_- q^+_+ + q^+_+ \int^x \left(q^+_+ q^-_- - \frac{U_+ U_-}{q^+_+ q^-_-}\right) dx' = 0, \\
\partial_+ q^-_- + q^-_- \int^x \left(q^+_+ q^-_- - \frac{U_+ U_-}{q^+_+ q^-_-}\right)  dx' = 0, \\
\partial_+ U_- = 0,\,\,\,\partial_- U_+ = 0.
\end{eqnarray}
We notice that both models, considered in this section and the previous one, are non-local. However,
in the next section we construct the model with local interaction term of four fermions type.

 \section{Majorana-Thirring type model}

More general type of nonlinear $\sigma$ model corresponds to time independent Landau-Lifshitz equation (\ref{TM1}) in moving frame 
\begin{equation}
v^\mu \partial_\mu {\bf s} = {\bf s} \wedge \partial^\mu \partial_\mu {\bf s},
\end{equation}
with constant vector ${\bf v} = (v^0, v^1)$ and $diag(1,-1)$ psudo-Euclidean metric. This corresponds to 
non-compact one sheet hyperbolic version of model \cite{Papa}. In terms of the light cone variables $v^+ = \frac{1}{2}(v^0 + v^1)$, $v^- = \frac{1}{2}(v^0 - v^1)$
we have equation
\begin{equation}v^{+}\partial_{+}{\bf s} +
  v^{-}\partial_{-}{\bf s}
= {\bf s}\wedge \partial_{+}\partial_{-}{\bf s}, \label{NPML}\end{equation}
where $\partial_{\pm} = \partial_{0} \pm \partial_{1} $.
This model produces constraints
\begin{eqnarray} D^{-}_{+}q^{+}_{-} &=& v^{+}q^{+}_{+} + v^{-}q^{+}_{-}, \label{nconstr1} \\
D^{+}_{+}q^{-}_{-} &=& -v^{+}q^{-}_{+} - v^{-}q^{-}_{-}. \label{nconstr2}\end{eqnarray}
The system (\ref{ZC1}), (\ref{ZC2}), (\ref{ZC3}), with these gauge constraints  completely
characterizes the model. Indeed, combining them together we find
four equations
\begin{eqnarray}D^{-}_{-}q^{+}_{+} &=& v^{+}q^{+}_{+} + v^{-}q^{+}_{-},  \label{1}\\
D^{+}_{+}q^{-}_{-} &=& -v^{-}q^{-}_{-} - v^{+}q^{-}_{+}, \label{2}\\
D^{-}_{+}q^{+}_{-} &=& v^{-}q^{+}_{-} + v^{+}q^{+}_{+},  \label{3}\\
D^{+}_{-}q^{-}_{+} &= & -v^{+}q^{-}_{+} - v^{-}q^{-}_{-}, \label{4}\end{eqnarray}
which provide the conservation law
\begin{equation}\partial_{-} (v^{+} q^{+}_{+}q^{-}_{+})  +
\partial_{+} (v^{-} q^{+}_{-}q^{-}_{-})  = 0.
\end{equation}
Combining this equation with (\ref{ZC3}) we get the flatness condition
\begin{equation}\partial_{-}A_{+} - \partial_{+}A_{-} = 0,\end{equation}
for Abelian vector
potential
\begin{equation}A_{+} = V_{+} - {2\over v^{-}} q^{+}_{+}q^{-}_{+},\,\,
  A_{-} = V_{-} - {2\over v^{+}} q^{+}_{-}q^{-}_{-}.\label{Aflat}\end{equation}
In terms of the covariant derivatives,
$ {\cal D}^{\pm} = \partial \pm 1/2 A $ Eqs. (\ref{1})-(\ref{4}) become
\begin{eqnarray}{\cal D}^{-}_{-}q^{+}_{+} - v^{+}q^{+}_{+} - v^{-}q^{+}_{-}
- {1\over v^{+}}q^{+}_{+}q^{-}_{-}q^{+}_{-}& = &0, \\
{\cal D}^{+}_{+}q^{-}_{-} + v^{-}q^{-}_{-} + v^{+}q^{-}_{+}
+ {1\over v^{-}}q^{+}_{+}q^{-}_{-}q^{-}_{+} &=& 0, \\
{\cal D}^{-}_{+}q^{+}_{-} - v^{-}q^{+}_{-} - v^{+}q^{+}_{+}
- {1\over v^{-}}q^{+}_{-}q^{-}_{+}q^{+}_{+} &=& 0, \\
{\cal D}^{+}_{-}q^{-}_{+} + v^{+}q^{-}_{+} + v^{-}q^{-}_{-}
+ {1\over v^{+}}q{+}_{-}q^{-}_{+}q^{-}_{-} &= &0. \end{eqnarray}

By choosing constant value potentials
\begin{equation} A_{+} = -2v^{-},\,\,A_{-} = -2v^{+},\end{equation}
rescaling
\begin{equation}
q^+_+ = \sqrt{v^-} Q^+_+, \,\,\,q^-_+ = \frac{1}{\sqrt{v^+}} Q^-_+, \,\,\,q^+_- = \sqrt{v^+} Q^+_-, \,\,\,q^-_- = \frac{1}{\sqrt{v^-}} Q^-_-,
\end{equation}
and restricting velocity 
\begin{equation}  v^+ v^- =1,  \label{velocityconst} \end{equation}
so that
\begin{equation}q^{\pm}_{+} = \sqrt{v^{-}}Q^{\pm}_{+},\,\,
  q^{\pm}_{-} = \frac{1}{\sqrt{v^{-}}}Q^{\pm}_{-}, \label{rescale}\end{equation}
we obtain the system of nonlinear equations for  the real valued analog of Thirring model
\begin{eqnarray}{\partial}_{-}Q^{+}_{+} - Q^{+}_{-}
- Q^{+}_{+}Q^{-}_{-}Q^{+}_{-} &=& 0, \label{Q1}\\
{\partial}_{+}Q^{-}_{-} + Q^{-}_{+}
+ Q^{+}_{+}Q^{-}_{-}Q^{-}_{+} &= &0, \label{Q2}\\
{\partial}_{+}Q^{+}_{-} - Q^{+}_{+}
- Q^{+}_{-}Q^{-}_{+}Q^{+}_{+} &=& 0, \label{Q3}\\
{\partial}_{-}Q^{-}_{+} + Q^{-}_{-}
+ Q^{+}_{-}Q^{-}_{+}Q^{-}_{-} &=&0. \label{Q4}\end{eqnarray}

 The above procedure allows us to derive also the 
linear problem   corresponding  to this model.
The current (\ref{JMU}) in the light
cone variables  after
redefining the Abelian gauge potential by (\ref{Aflat}) and using (\ref{velocityconst}) gives
   the pair
\begin{equation}J_{\pm } = {1\over 2}\left(-v^{\mp} +
{1\over v^{\mp}} q^+_\pm q^-_\pm\right)\sigma_{3} +
\left(\matrix{0& q^-_{\pm} \cr q^+_{\pm}&0 \cr}\right).
\end{equation}
In terms of the rescaled fields (\ref{rescale}) we have the Lax pair (in zero-curvature condition form) for our model
\begin{eqnarray} J_{+} &=& {1\over 2}\left(-\lambda^{2} +
Q^+_+ Q^-_+\right)\sigma_{3} + \lambda
\left(\matrix{0& Q^-_{+} \cr Q^+_{+}&0 \cr}\right),\label{nlinear1}\\
J_{-} &=& {1\over 2}\left(- \frac{1}{\lambda^{2}} + Q^+_- Q^-_-\right)\sigma_{3} + {1\over \lambda}
\left(\matrix{0& Q^-_{-} \cr Q^+_{-}&0 \cr}\right),\label{nlinear2}
\end{eqnarray}
where the spectral parameter is
$v^{-} \equiv \lambda^{2} $.

\subsection{Hamiltonian structure}

It is convenient to change notations
\begin{equation}
Q^+_+ = p^+, \,\,Q^-_+ = p^-,\,\, Q^+_- = q^+,\,\, Q^-_- =q^-
\end{equation}
and represent system (\ref{Q1}) - (\ref{Q4}) in the form
\begin{eqnarray}
- \partial_- p^+ + q^+ + q^+ q^- p^+ &=& 0, \label{1p}\\
 \partial_- p^- + q^- + q^+ q^- p^-& = &0, \label{2p}\\
- \partial_+ q^+ + p^+ + p^+ p^- q^+ &= &0, \label{1q}\\
 \partial_+ q^- + p^- + p^+ p^- q^-& =& 0. \label{2q}
\end{eqnarray}
The system is Lagrangian with density 
\begin{equation}
L = - p^+ \partial_0 p^- - q^+ \partial_0 q^- + p^+ \partial_1 p^- - q^+ \partial_1 q^- - p^+ q^- - q^+ p^- - p^+ p^- q^+ q^-,
\end{equation}
and it is Hamiltonian with Hamiltonian functional
\begin{equation}
H = \int^\infty_{-\infty} (-p^+ \partial_1 p^- + q^+ \partial_1 q^- + p^+ q^- + q^+ p^- + p^+ p^- q^+ q^-) dx^1. \label{H}
\end{equation}
The corresponding Poisson brackets
\begin{equation}
\{ A, B\} = \int^\infty_{-\infty}\left(\frac{\partial A}{\partial p^+} \frac{\partial B}{\partial p^-} - \frac{\partial A}{\partial p^-} \frac{\partial B}{\partial p^+}
+ \frac{\partial A}{\partial q^+} \frac{\partial B}{\partial q^-} - \frac{\partial A}{\partial q^-} \frac{\partial B}{\partial q^+}\right) dx^1
\end{equation}
for canonical variables
\begin{eqnarray}
\{p^+(x^0,x^1), p^-(x^0,x'^1) \} =\delta(x^1-x'^1),\\
\{q^+(x^0,x^1), q^-(x^0,x'^1) \} =\delta(x^1-x'^1),
\end{eqnarray}
give Hamiltonian evolution equations 
\begin{equation}
\dot p^{\pm} = \{p^\pm, H\} = \pm \frac{\partial H}{\partial p^{\mp}},\,\,\,\,\dot q^{\pm} = \{q^\pm, H\} = \pm \frac{\partial H}{\partial q^{\mp}}.
\end{equation}
Beside Hamiltonian (\ref{H}), exists integral of motion 
\begin{equation}
M = \int^\infty_{-\infty} (p^+ p^- + q^+ q^-)\, dx^1, \label{I}
\end{equation}
which plays role of the mass. In addition, one more conserved the momentum integral is
\begin{equation}
P = \int^\infty_{-\infty} (p^+ \partial_1 p^- + q^+ \partial_1 q^-)\, dx^1.\label{P}
\end{equation}
These three intergrals are the first ones of an infinite set of integrals of motion, which can be calculated from the linear problem.

\subsection{Hyperbolic complex Thirring form}

Here, by introducing the hyperbolic  complex variables or the "double numbers" \cite{Yaglom},  we represent our main system (\ref{1p}) - (\ref{2q}) in form of the hyperbolic complex Thirring type model.
By introducing four real functions
\begin{equation}
q^\pm = u_1 \pm v_1,\,\,\,\,p^\pm = u_2 \pm v_2,
\end{equation} 
the system can be rewritten as
\begin{eqnarray}
- \partial_+ v_1 + u_2  + (u^2_2 - v^2_2)  u_1  &=& 0, \label{1u}\\
 -\partial_+ u_1 + v_2  + (u^2_2 - v^2_2)  v_1  &=& 0, \label{2u}\\
-\partial_- v_2 + u_1  + (u^2_1 - v^2_1)  u_2  &=& 0, \label{1v}\\
-\partial_- u_2 + v_1  + (u^2_1 - v^2_1)  v_2  &=& 0 . \label{2v}
\end{eqnarray}
Now we combine these functions as the hyperbolic complex valued functions (or double number valued functions)
\begin{equation}
\chi_1 = u_1 + j v_1,\,\,\,\,\chi_2 = u_2 + j v_2,
\end{equation}
and corresponding conjugate functions
\begin{equation}
\bar \chi_1 = u_1 - j v_1,\,\,\,\,\bar\chi_2 = u_2 - j v_2,
\end{equation}
so that
\begin{equation}
\bar \chi_1 \chi_1 = |\chi_1|^2 = u^2_1 - v^2_1,\,\,\,\,\bar \chi_2 \chi_2 = |\chi_2|^2 = u^2_2 - v^2_2,
\end{equation}
where hyperbolic imaginary unit satisfies
\begin{equation}
j^2 =1, \,\,\,\,\bar j = - j.
\end{equation}
In matrix representation this unit can be defined as $j = \sigma_1$. In terms of these functions, our model takes the form
\begin{eqnarray}
- j\partial_+ \chi_1 + \chi_2  + |\chi_2|^2  \chi_1  &=& 0, \label{h1}\\
 -j \partial_- \chi_2 + \chi_1  + |\chi_1|^2  \chi_2  &=& 0, \label{h2} 
\end{eqnarray}
of the hyperbolic complex Thirring model. This representation is remarkable since the equation formally looks similar to the
usual Thirring model for complex functions $\psi_1$, $\psi_2$ and hyperbolic imaginary unit $j$ replaced by usual complex unit $i = \sqrt{-1}$.

\subsection{Dynamical System}

For homogeneous configurations $\partial_1 =0$, we get four dimensional
dynamical system
\begin{eqnarray}
\dot X_{1} &=& X_{3} + X_{1}X_{3}X_{4}, \\
\dot X_{2} &=& -X_{4} - X_{2}X_{3}X_{4}, \\
\dot X_{3} &=& X_{1} + X_{3}X_{1}X_{2}, \\
\dot X_{4} &=& -X_{2} - X_{4}X_{1}X_{2}, 
\end{eqnarray}
with the first integral
\begin{equation}I = X_{1}X_{2} + X_{3}X_{4} = const, \label{I1}\end{equation}
where
\begin{eqnarray}X_{1} \equiv Q^{+}_{+}
,\,\,X_{2} \equiv Q^{-}_{+}
,\,\,X_{3} \equiv Q^{+}_{-}
,\,\,X_{4} \equiv Q^{-}_{-}. \end{eqnarray}
The system is Hamiltonian with the canonical pairs
\begin{equation}\{X_{1},X_{2}\} = 1,\,\, \{X_{3},X_{4}\} = 1 ,\end{equation}
and the Hamiltonian function
\begin{equation}H = X_{2}X_{3} + X_{1}X_{4} + X_{1}X_{2}X_{3}X_{4}.
\end{equation}
The Hamiltonian provides the second integral of the motion,
and as easy to check the  integrals $I$ and $H$ are in involution,
$\{I,H\} = 0$. Integral (\ref{I1}) generates
the scaling transformation
$$\delta X_{i} = \{X_{i},I\}\alpha = \alpha X_{i},
(i = 1,3), \,\,\,
\delta X_{j} = \{X_{j},I\}\alpha = -\alpha X_{j},
(j = 2,4),$$
or after integration
$$X'_{1} = e^{\alpha}X_{1},\,\,
X'_{2} = e^{-\alpha}X_{2},\,\,
X'_{3} = e^{\alpha}X_{3},\,\,
X'_{4} = e^{-\alpha}X_{4}.$$

\section{Bilinear Form and Dissipaton solution}

The bilinear form for system (\ref{1p})-(\ref{2q}) can be derived in terms of six real functions, $g^\pm, h^\pm, f^\pm$, so that 
\begin{equation}
p^\pm = \frac{g^\pm}{f^\mp} = \frac{g^\pm f^\pm}{f^\pm f^\mp},\,\,\,\,q^\pm = \frac{h^\pm}{f^\pm} = \frac{h^\pm f^\mp}{f^\mp f^\pm}
\end{equation}
by representing system in the form
\begin{equation}
\mp \frac{D_-(g^\pm \cdot f^\pm)}{f^\pm f^\mp} \mp \frac{g^\pm}{f^\pm}\frac{D_-(f^\pm \cdot f^\mp)}{(f^\mp)^2} + \frac{h^\pm f^\mp}{f^\pm f^\mp} + \frac{h^+ h^-}{f^+ f^-} 
\frac{g^\pm}{f^\mp} = 0,
\end{equation}
\begin{equation}
\mp \frac{D_+(h^\pm \cdot f^\mp)}{f^\mp f^\pm} \mp \frac{h^\pm}{f^\mp}\frac{D_+(f^\mp \cdot f^\pm)}{(f^\pm)^2} + \frac{g^\pm f^\pm}{f^\mp f^\pm} + \frac{g^+ g^-}{f^- f^+} 
\frac{h^\pm}{f^\pm} = 0.
\end{equation}
This can be splited to bilinear system of equations
\begin{eqnarray}
\mp D_- (g^\pm \cdot f^\pm) + h^\pm f^\mp &=& 0, \\
\mp D_+ (h^\pm \cdot f^\mp) + g^\pm f^\pm &=& 0, \\
 D_+ (f^+ \cdot f^-) + g^+ g^- &=& 0, \\
-D_- (f^+ \cdot f^-) + h^+ h^- &=& 0.
\end{eqnarray}
Let $x^0 \equiv T$, $x^1 \equiv X$ are time and space coordinates in laboratory coordinate systems, and 
\begin{equation}
x = \frac{1}{2}(X+T),\,\,\,\,t = \frac{1}{2} (X-T),
\end{equation}
are the light-cone coordinates, so that
$X = x+t$, $T = x-t$, and
\begin{eqnarray}
\partial_- &=& \frac{\partial}{\partial x^0} - \frac{\partial}{\partial x^1} = \frac{\partial}{\partial T} - \frac{\partial}{\partial X} = -\frac{\partial}{\partial t},\\
\partial_+ &=& \frac{\partial}{\partial x^0} + \frac{\partial}{\partial x^1} = \frac{\partial}{\partial T} + \frac{\partial}{\partial X} = \frac{\partial}{\partial x}.
\end{eqnarray}
By rewriting Hirota derivatives in light-cone coordinates, $D_- = - D_t$, $D_+ = D_x$ the bilinear system becomes
\begin{eqnarray}
\pm D_t (g^\pm \cdot f^\pm) + h^\pm f^\mp &=& 0, \\
\mp D_x (h^\pm \cdot f^\mp) + g^\pm f^\pm &=& 0, \\
 D_x (f^+ \cdot f^-) + g^+ g^- &=& 0, \\
D_t (f^+ \cdot f^-) + h^+ h^- &=& 0.
\end{eqnarray}
so that 
\begin{equation}
q^+ q^-(x,t) = - \left( \ln \frac{f^+}{f^-}\right)_t, \,\,\,\, p^+ p^-(x,t) = - \left( \ln \frac{f^+}{f^-}\right)_x.
\end{equation}
By Hirota expansion 
\begin{eqnarray}
g^\pm(x,t) &=& \epsilon g_1^\pm(x,t) + \epsilon^3 g^\pm_3(x,t) +...\\
h^\pm(x,t) &=& \epsilon h_1^\pm(x,t) + \epsilon^3 h^\pm_3(x,t) +...\\
f^\pm(x,t) &=& 1 + \epsilon^2 f^\pm_2(x,t) +...,\,\,\,
\end{eqnarray} 
we find exact solution in the form
\begin{eqnarray}
g^\pm_1 &=& e^{\eta^\pm_1},\,\,\,\,\, h^\pm_1 = a^\pm_1 e^{\eta^\pm_1},\,\,\,\,\, f^\pm_2 = b^\pm_2 e^{\eta^+_1 + \eta^-_1}, \\
b^+_2 &=& \frac{(a^+_1)^2 a^-_1}{(a^+_1 - a^-_1)^2},\,\,\,\,b^-_2 = \frac{a^+_1 (a^-_1)^2}{(a^+_1 - a^-_1)^2}, \\
\eta^\pm_1 &=& k^\pm_1 x + \omega^\pm_1 t + \eta^\pm_{1_0}, \,\,\,\omega^\pm_1 = \mp a^\pm_1, \,\,\,k^\pm_1 = \pm \frac{1}{a^\pm_1},
\end{eqnarray}
parametrized by real constants $a^\pm_1$, $\eta^\pm_{1_0}$, so that 
\begin{equation}
\eta^\pm_1 = \pm \left( \frac{1}{a^\pm_1} x - a^\pm_1 t           \right) + \eta^\pm_{1_0}.
\end{equation}
This gives dissipative one-soliton solution, known as  the dissipaton, in the following form
\begin{eqnarray}
p^\pm (x,t) = \frac{g^\pm}{f^{\mp}} = \frac{e^{\eta^{\pm_1}}}{1 + b_2^{\mp} e^{\eta^+_1 + \eta^-_1}},\\
q^\pm (x,t) = \frac{h^\pm}{f^{\pm}} = \frac{a^{\pm_1} e^{\eta^{\pm_1}}}{1 + b_2^{\pm} e^{\eta^+_1 + \eta^-_1}}.
\end{eqnarray}
Components of this solution are decaying and growing exponentially
\begin{eqnarray}
p^\pm(x,t) &=& \frac{e^{\pm \frac{\eta^+_1-\eta^-_1}{2} - \frac{\alpha_\mp}{2}}}{2 \cosh \frac{\eta^+_1 + \eta^-_1 + \alpha_{\mp}}{2}},\\
q^\pm(x,t) &=& \frac{a^\pm_1 e^{\pm \frac{\eta^+_1-\eta^-_1}{2} - \frac{\alpha_\pm}{2}}}{2 \cosh \frac{\eta^+_1 + \eta^-_1 + \alpha_{\pm}}{2}},
\end{eqnarray}
where $\alpha_\pm = \frac{1}{2} \ln b_2^\pm$. But the mutual products 
are in perfect soliton form 
\begin{eqnarray}
p^+p^- &=& \frac{(a^+_1 -a^-_1)^2}{a^+_1 a^-_1} \frac{1}{2 \sqrt{a^+_1 a^-_1} \cosh\left( \eta^+_1 + \eta^-_1 + \frac{\alpha_+ + \alpha_-}{2}\right) + a^+_1 + a^-_1},\\
q^+q^- &=& {(a^+_1 -a^-_1)^2} \frac{1}{2 \sqrt{a^+_1 a^-_1} \cosh\left( \eta^+_1 + \eta^-_1 + \frac{\alpha_+ + \alpha_-}{2}\right) + a^+_1 + a^-_1}.
\end{eqnarray}
To have non-singular solution we choose real parameters $a^+_1 > 0$, $a^-_1 >0$ and as follows $b^\pm_2 >0$. This implies $k^+_1 >0$ and $k^-_1 < 0$.
By introducing parametrization
\begin{equation}
a^+_1 = \lambda_1 + \mu_1,\,\,\,a^-_1 = \lambda_1 - \mu_1,
\end{equation}
so that $a^+_1 a^-_1 = \lambda^2_1 - \mu^2_1$, $a^+_1 + a^-_1 = 2 \lambda_1$, $a^+_1 - a^-_1 = 2 \mu_1$ we have
\begin{eqnarray}
p^+p^- &=& \frac{1}{\lambda^2_1 - \mu^2_1} \frac{2 \mu^2_1}{\sqrt{\lambda^2_1 -\mu^2_1} \cosh\left( \eta^+_1 + \eta^-_1 + \frac{\alpha_+ + \alpha_-}{2}\right) + \lambda_1},\\
q^+q^- &=& \frac{2 \mu^2_1}{\sqrt{\lambda^2_1 -\mu^2_1} \cosh\left( \eta^+_1 + \eta^-_1 + \frac{\alpha_+ + \alpha_-}{2}\right) + \lambda_1}.
\end{eqnarray}
The traveling wave factor in these expressions can be rewritten in the laboratory coordinates $(X,T)$ with relativistic Lorentz contraction factor
\begin{equation}  
\eta^+_1 + \eta^-_1 + \frac{\alpha_+ + \alpha_-}{2} = - 2 k \frac{X-X_0 - v T}{\sqrt{1 - v^2}},
\end{equation}
where velocity of the dissipaton is defined as
\begin{equation}
v \equiv \frac{a^+_1 a^-_1 - 1}{a^+_1 a^-_1 + 1} = \frac{\lambda^2_1 - \mu^2_1 - 1}{\lambda^2_1 - \mu^2_1 + 1}
\end{equation}
and it is restricted by the speed of light $c=1$: $|v| < 1$. The initial position is fixed by
\begin{equation}
\frac{2k}{\sqrt{1-v^2}} X_0 \equiv \eta^+_{1_0} + \eta^-_{1_0} + \ln \frac{1}{4 k^2} \sqrt{\frac{1+v}{1-v}}
\end{equation}
and 
\begin{equation}
k \equiv \mu_1 \sqrt{\frac{1-v}{1+v}}.
\end{equation}
In terms of parameters $v$, $k$ and $X_0$, dissipaton densities become just
\begin{eqnarray}
p^+p^- = \sqrt{\frac{1-v}{1+v}} \frac{2 k^2 }{\cosh\left( 2k \frac{X-X_0 - v T}{\sqrt{1-v^2}}\right) + \sqrt{k^2 +1}},\label{densityp}\\
q^+q^- = \sqrt{\frac{1+v}{1-v}} \frac{2 k^2 }{\cosh\left( 2k \frac{X-X_0 - v T}{\sqrt{1-v^2}}\right) + \sqrt{k^2 +1}}.\label{densityq}
\end{eqnarray}
By expressing 
\begin{equation} 
\alpha_\pm = \ln \left(\frac{\sqrt{k^2 +1} \pm k}{4 k^2} \sqrt{\frac{1+v}{1-v}}\right),
\end{equation}
\begin{equation}
k \frac{X_{0_\pm}}{\sqrt{1-v^2}} = \frac{\eta^+_{1_0} + \eta^-_{1_0}}{2} + \frac{1}{2} \alpha_\pm,
\end{equation}
and denoting $\eta^+_{1_0} - \eta^-_{1_0} \equiv \nu_{1_0}$, we finally get dissipaton solution in the form
\begin{eqnarray}
p^\pm = \left( \frac{1-v}{1+v}\right)^{\frac{1}{4}} \frac{k \sqrt{\sqrt{k^2 +1} \pm k}}{\cosh k \frac{X - X_{0_\mp} -v T}{\sqrt{1-v^2}}} 
e^{\pm \left[ \frac{\sqrt{k^2 +1}}{\sqrt{1-v^2}}  (T - v X) + \nu_{1_0}  \right]}, \label{Dp}\\
q^\pm = \left( \frac{1+v}{1-v}\right)^{\frac{1}{4}} \frac{k \sqrt{\sqrt{k^2 +1} \pm k}}{\cosh k \frac{X - X_{0_\pm} -v T}{\sqrt{1-v^2}}} 
e^{\pm \left[ \frac{\sqrt{k^2 +1}}{\sqrt{1-v^2}}  (T - v X) + \nu_{1_0}  \right]}.\label{Dq}
\end{eqnarray}
It is noticed that initial positions are related by mean value formula
\begin{equation}
X_0 = \frac{1}{2} (X_{0_+} + X_{0_-}).
\end{equation}

\section{Relativistic Dissipaton}

For physical interpretation we calculate the mass (\ref{I}), momentum (\ref{P}) and energy integrals (\ref{H}), corresponding to one dissipaton solution (\ref{Dp}), (\ref{Dq}).
Substituting densities (\ref{densityp}), (\ref{densityq}) to (\ref{I}), after integration we get the mass integral as function of $k$ only,
\begin{equation}
M = 2 \ln  \frac{\sqrt{k^2 +1} + |k|}{\sqrt{k^2 +1} - |k|}. \label{mass}
\end{equation}
The momentum integral (\ref{P}) on one dissipaton solution takes the form
\begin{equation}
P = \frac{4 k v}{\sqrt{1-v^2}}.
\end{equation}
To calculate the energy, we first rewrite (\ref{H}) in the form
\begin{eqnarray}
H = \int^\infty_{-\infty} [ \frac{1}{2}(-p^+ \partial_1 p^- + p^- \partial_1 p^+ + q^+ \partial_1 q^- - q^- \partial_1 q^+)  \\  + p^+ q^- + q^+ p^- + p^+ p^- q^+ q^- ]\, dX =\\
\int^\infty_{-\infty} [\frac{1}{2} p^+ p^- \partial_1 \left( \ln \frac{g^+}{g^-}\right) - \frac{1}{2} q^+ q^- \partial_1\left( \ln \frac{h^+}{h^-}\right) + \\
+ p^+ q^- + p^- q^+ -\frac{1}{4} (p^+ p^- - q^+ q^-)^2 ]\, dX.
\end{eqnarray}
For one dissipaton solution it gives 
\begin{equation}
E = \frac{4k}{\sqrt{1-v^2}}.
\end{equation}
Denoting 
\begin{equation}
m_0 \equiv 4 k
\end{equation}
as the rest mass, we find usual expressions for momentum and energy of relativisitic particle
\begin{equation}
P = \frac{m_0 v}{\sqrt{1-v^2}},\,\,\,\,E = \frac{m_0}{\sqrt{1-v^2}},
\end{equation}
with speed of light $c=1$.
This shows that our one-dissipaton solution describes a finite energy relativisitc particle with the rest mass $m_0$, corresponding to the rest frame, when 
$v=0$. The dispersion relation for one-dissipaton is in the relativistic form
\begin{equation}
E^2 - P^2 = m_0^2.
\end{equation}
However, the rest mass $m_0$ is connected with first integral $M$ by nonlinear formula
\begin{equation} 
M = 2 \ln \frac{\sqrt{m^2_0 + 16} + m_0}{\sqrt{m_0^2 + 16} - m_0}
\end{equation}
or 
\begin{equation}
m_0 = 4 \sinh \frac{M}{4}.
\end{equation}
It shows that in contrast with known non-relativisitc dissipatons of RNLS \cite{PL2002}, our dissipaton is a composite object with properties of relativisitc particle.

\subsection{Resonant interaction}

It is well known that non-relativistic dissipatons, related to  RNLS and RDNLS equations show resonant properties under collisions\cite{PL2002}, \cite{PLANZIAM2002}. An interesting point is to see if such
property is preserved also in the relativistic case. To explore this possibility, we consider collision of two dissipatons with masses and velocities $(m_1, v_1)$ and $(m_2, v_2)$,
producing by  fusion one dissipaton $(m,v)$.The conservation laws for this process implies following relations
\begin{equation}
M = M_1 + M_2,\,\,\,P = P_1 + P_2,\,\,\,\,E= E_1 + E_2.
\end{equation}
Substituting to above formulas we have equations for resonant interaction of dissipatons
\begin{eqnarray} 
\frac{\sqrt{m^2_1 + 16} + m_1}{\sqrt{m_1^2 + 16} - m_1} \,\frac{\sqrt{m^2_2 + 16} + m_2}{\sqrt{m_2^2 + 16} - m_2} &=& \frac{\sqrt{m^2 + 16} + m}{\sqrt{m^2 + 16} - m}, \label{R1}\\
\frac{m_1 v_1}{\sqrt{1-v_1^2}} + \frac{m_2 v_2}{\sqrt{1-v_2^2}} &=& \frac{m v}{\sqrt{1-v^2}},\label{R2}\\
\frac{m_1}{\sqrt{1-v_1^2}} + \frac{m_2}{\sqrt{1-v_2^2}}& =& \frac{m}{\sqrt{1-v^2}}\label{R3}.
\end{eqnarray}
If this system of algebraic equations admits nontrivial solution, then interaction of our relativistic dissipatons could have resonant character.
We postpone the study of general solution for this system and will consider here only special case, namely collision of two equal mass dissipatons $m_1 = m_2$, with equal
and opposite velocities $v_1 = - v_2$. It implies   $P_1 = - P_2$  $\rightarrow$  $P_1 + P_2 = 0$ $\rightarrow$ $P=0$.  This process  creates one dissipaton with mass $m$ in the rest with $v = 0$. From (\ref{R1}) and (\ref{R3}) we have
\begin{eqnarray}
\left(\frac{\sqrt{m^2_1 + 16} + m_1}{\sqrt{m_1^2 + 16} - m_1}\right)^2  & =& \frac{\sqrt{m^2 + 16} + m}{\sqrt{m^2 + 16} - m} , \\
m &=& \frac{2 m_1}{\sqrt{1 - v_1^2}}.
\end{eqnarray}
Solution of this system is given by relation between velocity and mass of colliding dissipatons
\begin{equation}
v_1 = \frac{m_1}{\sqrt{m_1^2 + 16}},
\end{equation}
so that $v_1 < 1$, and mass of dissipaton in the rest is
\begin{equation}
m = \frac{1}{2} m_1 \sqrt{m_1^2 + 16}.
\end{equation}
It shows that similar to non-relativistic case, the  relativisitc dissipatons admit resonant interaction. 
Calculations of two dissipaton solution and study of their mutual resonant interaction for specific choice of parameters 
would be done in forthcoming publication.

\section{Conclusions} 
In present paper we have derived  new integrable nonlinear relativistic real valued spinor model with four fermionic interaction in Thirring type form. 
As was shown, the model is gauge equivalent to non-compact 
version of Papanicolau model  on one sheet hyperboloid and it provides a specific gauge constraint in JT gravity. By introducing bilinear form of the nonlinear system we 
calculated one dissipaton solution and corresponding integrals of motion as mass, momentum and energy. The obtained dispersion relation shows that dissipaton represents relativistic particle with 
highly nonlinear mass term. By analyzing resonans conditions for dissipaton scattering we found nontrivial solution with resonant properties. Description of this resonant
scattering requires calculation of two dissipaton solution in a specific range of parameters. Moreover, it is interesting to calculate one soliton solution of the spin model, corresponding to
dissipaton solution and the metric tensor in JT  gravity on existence of relativisitc black holes. This work is in progress now.

\section{Acknowledgments} This is a good chance to contribute a paper in this special issue of SEABM in honor of Ky Fan. Professor Ky Fan has been the Director of Institute of Mathematics, Academia Sinica,Taipei in the period of 1978-1984. He emphasized the importance of basic research and the need for the young research assistants to be exposed to more mathematics.
 One of the authors J.L entered this Institute in 1983  when Prof. Ky Fan was the Director.

\end{document}